\documentclass[a4paper, 10 pt, journal, twoside]{IEEEtran}  

\IEEEoverridecommandlockouts                              

 \usepackage{graphicx} 
\usepackage{times} 
\usepackage{amsmath} 
\usepackage{amssymb}  
\usepackage{color}
\usepackage{algorithmic}
\usepackage{flushend}

\newcommand{\R}{{\mathbb{R}}}

\newcommand{\C}{{\mathbb{C}}}

\newcommand{\Ecal}{{\cal E}}

\newcommand{\Hcal}{{\cal H}}

\newcommand{\Kcal}{{\cal K}}

\newcommand{\Scal}{{\cal S}}
\newcommand{\Ucal}{{\cal U}}
\newcommand{\Vcal}{{\cal V}}

\newtheorem{theorem}{Theorem}

\newtheorem{remark}{Remark}

\newtheorem{problem}{Problem}

\newtheorem{notation}{Notation}
\newtheorem{sam}{Example}

\title{
	``Weak'' Control for Human-in-the-loop Systems
}

\author{
 	Masaki Inoue and
	Vijay Gupta
 \thanks{M.~Inoue is with 
 			Department of Applied Physics and Physico-Informatics, Keio University, 
 		3-14-1 Hiyoshi, Kohoku-ku, Yokohama, Kanagawa 223-8522, Japan.
         {\tt\small minoue@appi.keio.ac.jp}}%
 \thanks{V.~Gupta is with 
 		Department of Electrical Engineering,
		University of Notre Dame,
 		275 Fitzpatrick Hl Engrng, Notre Dame, IN 46556, USA.
         {\tt\small vgupta2@nd.edu}}%
\thanks{*This work was supported by CREST No.~JPMJCR15K1 from JST and also by the Grant-in-Aid for Young Scientists (B), No.~17K14704 from JSPS.}%
 }
\begin{document}

\maketitle
\thispagestyle{empty}
\pagestyle{empty}

\begin{abstract}
In this letter, we propose a control framework for human-in-the-loop systems, in which 
many human decision makers are involved in the feedback loop composed of a plant and a controller.  
The novelty of the framework is that the decision makers are {\it weakly controlled};
in other words, they receive a set of admissible control actions from the controller 
and choose one of them in accordance with their private preferences. 
For example, the decision makers can decide their actions to minimize their own costs or by simply relying on their experience and intuition.  
A class of controllers which output set-valued signals is proposed, and it is shown that the overall control system is stable independently of the decisions
made by the humans.
Finally, a learning algorithm is applied to the controller that updates the controller parameters  to reduce the achievable minimal costs for the decision makers.  
Effective use of the algorithm is demonstrated in a numerical experiment.
\end{abstract}

\begin{IEEEkeywords}
Human-in-the-loop system, stability, optimization, internal model control, robust control
\end{IEEEkeywords}
\section{INTRODUCTION}
\IEEEPARstart{T}{his} letter is devoted to constructing a control framework for human-in-the-loop (HIL) systems, in which 
multiple decision makers are involved in the feedback loop composed of a plant  and a controller. 

In the last five decades, the HIL concept has been realized and developed significantly in the literature.  
Most works focus on cooperative operation of the human and autonomous plants such as robots.  
There have been a variety of frameworks for the analysis and design of such human-robots interaction 
(see, e.g. the pioneering works and survey papers \cite{Whitney_69,Mcruer_80,Sheridan_89,Hokayem_06,Goodrich_08} and 
recent trials \cite{Cao_08,Chipalkatty_13,Lam_14,Hatanaka_CDC15,Music_17}).

Applications of HIL systems are now being proposed beyond such human-robots systems,   
where cooperation between human and robot is the key.
Potential applications of HIL systems include 
for example, demand response in power grids involving humans decisions \cite{Palensky_11}, 
air traffic management that must include human factors for pilots and control centers \cite{Wickens_97},
incentive-based control of intelligent transportation systems relying on humans smart decisions \cite{Barfield_97}, and so on.  
In such systems, the priorities of the humans in the loop may be unknown to and misaligned with those of the system designer.
To  realize such systems and to further broaden the applications, 
a broader control framework for HIL systems is necessary.

Some works have tried to construct more general control frameworks for HIL systems 
in e.g. \cite{Chipalkatty_13,Lam_14,Hatanaka_CDC15,Maestre_14,Van_15,Feng_16,Eichler_17}.
In \cite{Feng_16,Eichler_17}, humans are modeled as uncertainties or constraints, 
and various methods of compensating their negative actions are proposed.
In \cite{Chipalkatty_13,Lam_14,Hatanaka_CDC15,Maestre_14,Van_15}, 
humans are positively involved  in the feedback loop of the controlled systems. 
In \cite{Chipalkatty_13,Lam_14,Hatanaka_CDC15}, humans are modeled as reference generators for autonomous controlled robots.  
This can be viewed as human decision-making being involved in the outer feedback loop of the overall control system.  The cooperation of the human and inner controller is achieved by model predictive control (MPC) scheme or passivity-property.
In the problem setting of \cite{Maestre_14,Van_15}, humans are involved in the inner feedback loop.  
In particular, humans handle both actuation and measurement of the plant based on the request by the controller.  
Humans are characterized by the intermittency of their control actions or measurements and their spatial mobility.  Then, an MPC-based method is proposed and applied to the practical control problem of an irrigation canal system.

In this letter, we propose a novel control framework for the HIL systems. 
In the framework, humans are interpreted as decision makers and are involved in the inner feedback loop of a plant and a controller.  
The humans handle the actuation to plant based on the request by the controller.  
We aim to realize ``weak control'' of the HIL system; 
the controller does not impose ``too severe'' requests for the decision makers that completely consume the degree of freedom (DOF) of their decisions.  
Instead, the controller provides a set of admissible control actions to enable the decision makers to pursue their own aims by utilizing the remaining DOF.

\begin{figure}[t]
 \begin{center}
 \includegraphics[width=0.8\hsize]{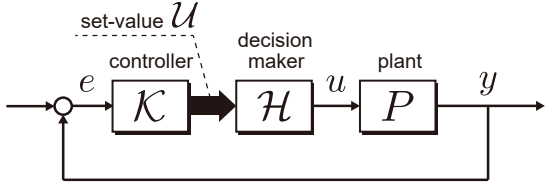}
 \caption{A framework of weak control for human-in-the-loop system.  The overall control system is composed of the plant $P$, controller $\Kcal$, and decision maker (e.g. humans) $\Hcal$.  A set-valued signal $\Ucal$ is generated by $\Kcal$ and is provided to $\Hcal$, and a signal $u$ is chosen as $u \in \Ucal$ by $\Hcal$ to actuate $P$.}%
 \label{fig:HIL}
 \end{center}
 \end{figure}

In the rest of the letter, 
first, the problem of the weak control for the HIL systems is formulated, 
in which the decision makers choose one control action $u$ from a given set of admissible actions $\Ucal$ as illustrated in Fig.~\ref{fig:HIL}. 
  Then, the solution is derived based on the idea of the internal model control (IMC, \cite{Garcia_82}).  The resulting controller
generates a set-valued signal, and it is shown that the overall control system is stable independently of the decisions.
Finally, a learning algorithm is applied to the controller that updates the controller parameters in order to reduce the achievable cost for the decision makers.  
Effective use of the framework is demonstrated in a numerical experiment of an HIL control problem.

Notation: 
Let $v$ and $\Vcal$ be a signal and set-valued signal, respectively.  Then,
their sum is defined as $v + \Vcal:= \{v + \tilde{v}\,|\, \tilde{v} \in \Vcal\}$. 
The symbol $I$ represents the identity operator, i.e.,  for any signal $u$, $I u = u$ holds.
For a given set $\Vcal$, the symbol $\Scal(\Vcal)$ represents 
an element of $\Vcal$, i.e., $\Scal(\Vcal)\in \Vcal$ holds. 
For a given input-output system $\Sigma$, 
the symbol $\| \Sigma \|$ represents some performance criterion of interest.

\section{HUMAN-IN-THE-LOOP CONTROL SYSTEMS}\label{sec:main}
\subsection{Problem Setting: Weak Control}
In this section, we formulate and solve the problem of  weak control for the HIL systems. 

\begin{figure}[t]
 \begin{center}
 \includegraphics[width=\hsize]{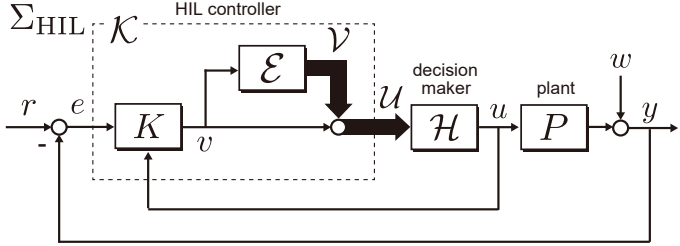}
 \caption{HIL controller and overall control system $\Sigma_{\rm HIL}$. }
 \label{fig:HIL2}
 \end{center}
 \end{figure}

The control structure for the HIL systems is illustrated in Fig.~\ref{fig:HIL2}, which is 
a specialization of the conceptual diagram illustrated in Fig.~\ref{fig:HIL}.  
In Fig.~\ref{fig:HIL2}, the plant $P$, decision maker $\Hcal$, and controller $\Kcal$ are connected to each other 
to construct the overall control system $\Sigma_{\rm HIL}$.  

The system description is given as follows.
The signals $r$ and $w$ are called
the reference and the disturbance, respectively.
The plant $P$ is a dynamical system that generates 
the output $y \in \R^\ell$ depending on the control input $u \in \R^m$.
The model of $P$ is described by 
\begin{align*}
	P:\ y = P(u) + w,
\end{align*}
where $P(\cdot)$ is an operator.  
The decision maker $\Hcal$ is a static system
 that generates $u(t)$ from a given input candidate $\Ucal(t) \subset \R^{m}$ for all $t$%
\footnote{It is assumed that the decision in $\Hcal$ is fast enough compared with the dynamic behavior of $P$. 
Therefore, $\Hcal$ is modeled as a static system in this letter. }.
The model of $\Hcal$ is described by 
\begin{align}
	\Hcal:\ u(t) = \Scal(\Ucal(t))\label{eq:H},
\end{align}
or equivalently by $\Hcal: u(t)  \in \Ucal(t)$.  The operator $\Scal$ represents the decision by $\Hcal$. 
The controller $\Kcal$ is a dynamical system that generates $\Ucal$ based on the error $e:=r -y$ and $u$.  
The controller $\Kcal$ is composed of an internal controller $K$ and an {\it expander} $\Ecal$.  The signal $v$ is generated by $K$ and is {\it expanded} to a set-valued signal $\Vcal$ by $\Ecal$.  The sum of $v$ and $\Vcal$ becomes the input candidate $\Ucal$.  The model of $\Kcal$ is described by
\begin{align*}
	{\cal K}:\
	\left\{\begin{array}{l}
	v = K(e, u), \\
	\Vcal = {\cal E}(v),\\
	{\cal U}= v + \Vcal, 
	\end{array}\right.
\end{align*}
where $K(\cdot,\cdot)$ and $\Ecal(\cdot)$ are operators.

The main characteristics of the proposed HIL system are the existence of a {\it set-valued} signal in the feedback loop. 
Due to this set-valued signal $\Ucal$, we say that the HIL system is {\it weakly controlled}.  
This weak control framework allows us to express the case that 
decision makers can freely choose their own actions to some extent.  
Thus, they can pursue their own benefits or simply rely on their experience and intuition for their choices. 
This freedom can be a useful feature in many problems involving humans in smart infrastructure systems, where the priorities of the humans may be private information or misaligned with those of the system operator, yet the system operator should give the human users sufficient freedom to choose from among a set of possible actions.

The HIL control problem addressed in this letter is summarized in the following problem.
\smallskip

\begin{problem} 
\label{prob}
{\it (HIL control problem):}\
Find $K$ and $\Ecal$ such that  $\Sigma_{\rm HIL}$ is input-output stable for all decisions by $\Hcal$.
\end{problem}
\smallskip

Note again that any strategy or model of $\Hcal$ is unavailable
for the  design of $K$ and $\Ecal$ in the general problem setting.
Only the rule (\ref{eq:H}) is known and available to the designer.

\subsection{Signal Expander}\label{sec:expand}
Examples of signal expanders $\Ecal$ are given in this subsection.

\begin{sam}
An example of the expander is given by the following {\it rectangular prism} $\Ecal_1$:
\begin{align*}
	\Ecal_1(v) =\left\{
\mbox{diag}(\delta_1,\ldots, \delta_m)
v\,\Bigg|\, \delta_i \in [\,-\gamma_i, \gamma_i\,]\right\},
\end{align*}
where $\gamma_i$, $i \in \{1,2,\ldots,m\}$ are positive constants.
Equivalently, this $\Ecal_1$ is written as 
\begin{align*}
	\Ecal_1(v) =\left\{
	\left[\begin{array}{c}
		\varepsilon_1 \\
		\vdots\\
		\varepsilon_m\\
	\end{array}\right]\,\Bigg|\, \varepsilon_i \in [\,-\gamma_i v_i, \gamma_i v_i\,]\right\}.
\end{align*}
\end{sam}

\begin{sam}
\label{ex:2}
The expander $\Ecal_1$ is generalized to $\Ecal_2$ with some coordinate transformation as:
\begin{align*}
	\Ecal_{\rm 2}(v) =
\left\{ E_{\rm L}
\mbox{diag}(\delta_1,\ldots, \delta_p)
E_{\rm R}^\top
v\, \Big|\, \delta_i \in [\,-\gamma_i, \gamma_i\,]\right\},
\end{align*}
where $p \leq m$ is a natural number, $E_{\rm L}\in \R^{m\times p}$ and $E_{\rm R}\in \R^{m\times p}$ are matrices of full column ranks.
By the introduction of $E_L$ and $E_R$, 
the signal $v$ is expanded more flexibly than $\Ecal_1(v)$.
Let us consider a simple example of $\Ecal_2$.  We define
\begin{align*}
	E_{\rm L} = E_{\rm R} = 	
	\frac{1}{\sqrt{2}}\left[\begin{array}{cc}
		1 \\
		-1 
	\end{array}\right].
\end{align*}
Then, $\Ecal_{\rm 2}(v)$ is reduced to
\begin{align*}
	\Ecal_{\rm 2}(v) 
	&= \left\{\frac{1}{2} 
	\left[\begin{array}{c}
		\varepsilon\\
		-\varepsilon 
	\end{array}\right]\Big|\, \varepsilon \in [\,-\gamma_2 |v_1 - v_2|, \gamma_2 |v_1 - v_2|\,]\right\}.
\end{align*}
We see that this $\Ecal_{\rm 2}(v) $ expands the signal $v$ such that the sum of the elements  is invariant.
\end{sam}

\begin{figure}[t]
 \begin{center}
 \includegraphics[width=0.85\hsize]{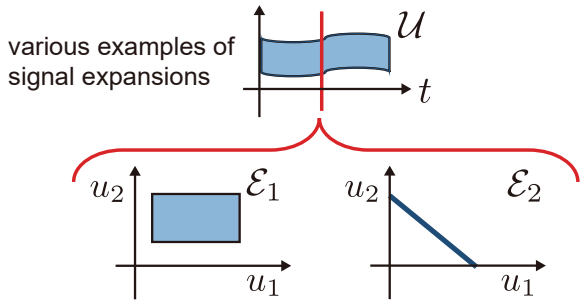}
 \caption{Examples of signal expansions.}%
 \label{fig:HILCinp}
 \end{center}
 \end{figure}

The set-valued signals $\Ucal$ generated by $\Ecal_{i}$, $i \in \{1,2\}$ are illustrated in Fig.~\ref{fig:HILCinp}.  Such generated $\Ucal$ must be a constraint for $\Hcal$ of  decision making.

\begin{remark}
Consider here that multiple decision makers $\Hcal_i$, $i \in \{1,2,\ldots,m\}$ are included in $\Hcal$ and they choose $u_i$, $i \in \{1,2,\ldots,m\}$, respectively by pursuing their own aims.
It should be noted that $\Ucal$ generated by $\Ecal_2$ implicitly requires {\it cooperation or negotiation}  between $\Hcal_i$, $i \in \{1,2,\ldots,m\}$ for their decision-making, 
while $\Ecal_1$ does not.
The decision makers $\Hcal_i$, $i \in \{1,2,\ldots,m\}$ must cooperate each other 
to determine their actions $u_i$, $i \in \{1,2,\ldots,m\}$ under  the {\it constraint} $u \in \Ucal$ for the case $\Ecal_2$.
\end{remark}

\subsection{Weak Control: IMC-based Approach}
\label{sec:sol}
In this subsection, we give a general solution to the HIL control problem, which is formulated in Problem \ref{prob}.

\begin{figure}[t]
 \begin{center}
 \includegraphics[width=0.6\hsize]{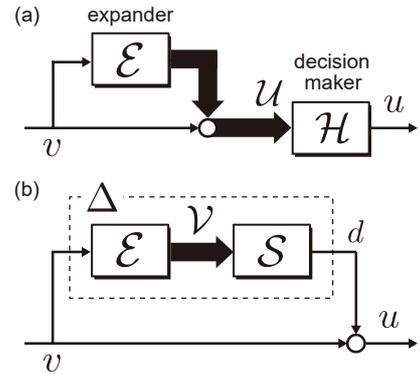}
 \caption{Transformation of expander $\Ecal$ and decision maker $\Hcal$.}%
 \label{fig:ContTrans}
 \end{center}
 \end{figure}

\begin{figure}[t]
 \begin{center}
 \includegraphics[width=\hsize]{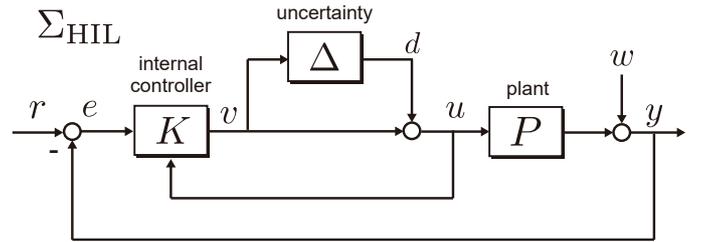}
 \caption{Transformation of overall control system $\Sigma_{\rm HIL}$.}%
 \label{fig:HIL_trans}
 \end{center}
 \end{figure}

First, the HIL control problem is reduced 
to a robust control problem \cite{Zhou_96} as follows.
Noting that $\Ucal = v + \Vcal$, the behavior of $\Hcal$ is equivalently expressed as
\begin{align*}
	\Hcal:\ 
	\left\{\begin{array}{l}
		d(t) = \Scal(\Vcal(t)),\\
		u(t)  = v(t) + d(t).
	\end{array}\right.
\end{align*}
This transformation is illustrated in Fig.~\ref{fig:ContTrans}.
Letting $\Delta$ be 
\begin{align}
	\Delta:\ d(t) = \Scal(\Ecal(v(t)))\label{eq:Delta}
\end{align}
or more simply $\Delta: d(t) \in \Ecal(v(t))$ as illustrated in Fig.~\ref{fig:ContTrans}(b),
we reduce the overall control system $\Sigma_{\rm HIL}$ to the system illustrated in Fig.~\ref{fig:HIL_trans}.  The system  illustrated in Fig.~\ref{fig:HIL_trans} represents a control system addressed in a robust control problem with the time-varying uncertainty $\Delta$.  

This transformation implies that the HIL control problem is essentially a robust control problem.  
Still, there are some practical differences between the problems considered in \cite{Zhou_96} and here.
The HIL control {\it positively} utilizes the uncertainty for the signal expansion, which brings some benefit to $\Hcal$.  On the other hand,
robust control focuses mainly on the {\it negative effect} of the uncertainty.
In addition, the uncertainty in the HIL control is {\it designable} to achieve some aims, while that in the robust control is not. 
Details of design examples and applications are given in 
Section \ref{sec:expand} and Section \ref{sec:app}.

Next, 
we derive a design method for $K$ based on the system illustrated in Fig.~\ref{fig:HIL_trans}.  
In particular, we propose a special controller-structure in $K$ to guarantee the stability of the overall control system $\Sigma_{\rm HIL}$ 
independently of the decisions made by $\Hcal$.

To this end, the internal controller $K$ is given by
\begin{align}
	v = K_e(e + P(u)),	\label{eq:cont}
\end{align}
where $K_e$ is an operator.
The controller structure in (\ref{eq:cont}), which involves the plant model $P$, 
is based on the idea of the internal model control (IMC, \cite{Garcia_82}).  
By applying (\ref{eq:cont}) to the system illustrated in Fig.~\ref{fig:HIL_trans}, we obtain the following theorem.
\medskip

\begin{theorem}
Suppose that $K(e,u)$ is given by (\ref{eq:cont}).  
Then, if $P$ and $K_e$ are $L_2$-stable,
$\Sigma_{\rm HIL}$  is $L_2$-stable for 
all decisions of $\Hcal$.
\end{theorem}
\medskip

{\it Proof:}
We recall that
\begin{align*}
	y &= P(u) +w,\\
	u &= v + \Delta(v),\\
	v & = K_e(r-y + P(u))
\end{align*}
hold, 
where $\Delta$ is the operator that represents the input-output map (\ref{eq:Delta}).
By summarizing the equations, we obtain the expression 
\begin{align}
	\Sigma_{\rm HIL}:\ y = P(K_e(r - w) + \Delta(K_e(r - w))) + w.\label{eq:HILall2}
\end{align}
From the cascaded and parallel structure in (\ref{eq:HILall2}), we see that the statement of the theorem holds.\hfill $\Box$
\smallskip

As stated in the proof of the theorem, the implementation of 
the IMC-based controller (\ref{eq:cont}) results in 
the cascaded and parallel structure in $\Sigma_{\rm HIL}$.
The structure contributes to the stability guarantee independently of the expander $\Ecal$ and the decision in $\Hcal$, which is described by $\Scal$ in (\ref{eq:Delta}).
In addition, the structure enables us to easily evaluate the performance of $\Sigma_{\rm HIL}$ as follows.
For simplicity, let us consider the linear regulation control problem; 
it is assumed that $r = 0$  and that $P$ and $K_e$ are linear.  
Then, the expression (\ref{eq:HILall2}) is reduced to 
\begin{align}
	\Sigma_{\rm HIL}:\ y = (I - P(I+\Delta)K_e)w.\label{eq:linearP}
\end{align}
Supposing $\Delta = 0$, i.e., $v$ is not expanded in $\Ecal$ or $u \equiv v$ is chosen in $\Hcal$,
we can evaluate the nominal performance $\| I - PK_e\|$ in some criterion such as the $L_2$ gain.
We emphasize that the performance $\| I - P(I+\Delta)K_e\|$ is {\it continuously and linearly} deteriorated from the nominal one with the increase of  $\|\Delta\|$. This enables us to simply evaluate the bound of $\| I - P(I+\Delta)K_e\|$.
The continuity of the performance deterioration is called {\it persistence} and analyzed for general uncertain systems in \cite{LCSS_18}.
\smallskip

\begin{remark}
\label{rem:design}
A design strategy of $K_e$ and $\Ecal$ is given in this remark.
First, we design $K_e$  such that the desired nominal performance is achieved; 
for example, minimize the performance as $\min \| I - PK_e\|=:\rho$.
Then, determine the degree of the expansion in $\Ecal$, which is characterized by e.g. $\gamma_i$  of $\Ecal_i$, $i\in \{1,2\}$.   
We design $\Ecal$ such that 
the performance deterioration is admissible for the designer, who is responsible for the overall control system; for example,  for a given $\Delta\rho >0$, find or maximize  $\gamma$ such that
\begin{align}
	\| I - P(I+\Delta)K_e\| \leq \rho + \Delta\rho
\label{eq:HILperformance}
\end{align}
 holds for all decisions in $\Hcal$ satisfying (\ref{eq:H}). 
\end{remark}

\section{LEARNING OF HUMAN PREFERENCES FOR UPDATING THE CONTROLLER}\label{sec:app}
In the general problem formulated in Section \ref{sec:main}, no assumption is imposed on the decision maker $\Hcal$ except for the rule (\ref{eq:H}).  
In this section, it is assumed that $\Hcal$ is rational and determines the control action $u$ based on an optimization;  a cost function is minimized under the constraint (\ref{eq:H}).
Then, we design and implement a mechanism of 
learning a part of the model in $\Hcal$ and 
of updating the expander $\Ecal$ online.

\subsection{Problem Setting}

The models of the plant $P$ and controller $\Kcal$ are specialized in the following discussion.
For simplicity, we consider a linear regulation problem under the step disturbance; 
 $r = 0$, $w$ is the step signal, $P$ and $K_e$ are linear, and the overall control system is expressed by (\ref{eq:linearP}).  
The following discussion can be extended to other practical cases, e.g.  
tracking control with $r\neq0$, persistent disturbance to $w$, nonlinear plant systems, and so on, with some modification.
In addition,
the structure of $\Ecal$ is fixed at $\Ecal_2$, which is defined in Example \ref{ex:2}.  
In addition, $\Ecal$ has only one dimensional degree of freedom; 
letting $E_{\rm L}$ and $E_{\rm R}$ be vectors in $\R^m$, $\Ecal$ is described by
\begin{align}
	\Ecal(v) = \Ecal_{\rm 2}(v) =
\left\{ \delta E_{\rm L} E_{\rm R}^\top
v\, \Big|\, \delta \in [\,-\gamma, \gamma\,]\right\},
\label{eq:specialE}
\end{align}
where $\gamma$ is a positive constant. 
Note here that $E_{\rm L}$ represents the {\it direction} of the expansion, while $\gamma |E_{\rm R}^\top v|$ represents the degree of the expansion.

We consider that the following optimization algorithm is implemented in $\Hcal$.
\begin{align} 
\Hcal:\
\left\{\begin{array}{ll}
\min\ &f(u)\\
\mbox{subject to}\ &u \in {\cal U}.
\end{array}\right.
\label{eq:Hopt}
\end{align}
The global minimizer of the {\it unconstrained} optimization, simply min $f(u)$,  is denoted by $u^\ast$, 
while that of the constrained one, described by (\ref{eq:Hopt}), is denoted by $u^\dagger$.
Trivially, $f(u^\ast) \leq f(u^\dagger)$ holds.
Note that 
 the achievable minimum cost $f(u^\dagger)$ 
depends on $\Ucal$, and therefore, it depends on the designed $\Ecal$. 
The aim of this section is to find $\Ecal$ that minimizes the achievable minimum cost $f(u^\dagger)$ subject to some performance specification on $\Sigma_{\rm HIL}$.

To formulate the problem in a clearer manner,
we define a specific set of expanders $\Ecal$, which is essentially the same as  a set of triplets $\{E_{\rm L}, E_{\rm R}, \gamma\}$, as follows.
Let $\rho$ be the nominal performance $\rho := \| I - PK_e\|$. 
\smallskip

\begin{notation}
For a given positive constant $\Delta\rho$, the symbol $\{\Ecal\}_{\Delta\rho}$ 
represents the set of the expanders $\Ecal$ 
such that
for any element in $\{\Ecal\}_{\Delta\rho}$, the inequality in (\ref{eq:HILperformance})
holds for all decisions by $\Hcal$, i.e., all realizations of $\Delta$.
In addition, 
$\{\Ucal(v)\}_{\Delta \rho}:= \{v + \Ecal(v)\,|\, \Ecal(v) \in \{\Ecal(v)\}_{\Delta \rho}\}$, which 
represents the set of all input candidates $\Ucal$ generated by $\Ecal(v) \in \{\Ecal(v)\}_{\Delta \rho}$. 
\end{notation}
\smallskip

The problem addressed in the rest of this section is formulated as follows as follows. 
\smallskip

\begin{problem}
\label{prob2}
For a given $\Delta\rho$, find $\Ecal \in \{\Ecal\}_{\Delta\rho}$ that minimizes 
 $f(u^\dagger)$ at the steady state.
\end{problem}
\smallskip

In the next subsection, the solution method by updating $\Ecal \in \{\Ecal\}_{\Delta\rho}$ is given.



\subsection{Learning Algorithm for Updating Expander}

\begin{figure}[t]
 \begin{center}
 \includegraphics[width=0.8\hsize]{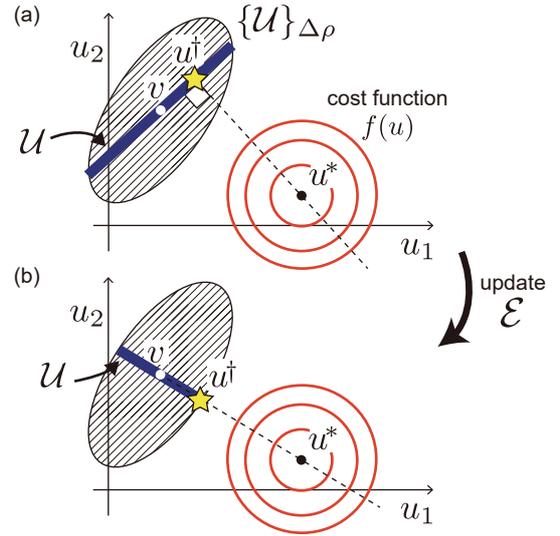}
 \caption{Graphical interpretation of the input candidate $\Ucal$ and decision by $\Hcal$.  
If $\Ucal \in \{\Ucal\}_{\Delta \rho}$ is provided by the controller $\Kcal$, rational $\Hcal$ chooses $u^\dagger$, which is the minimizer of the constrained optimization  problem (\ref{eq:Hopt}). 
}%
 \label{fig:U-cost}
 \end{center}
 \end{figure}

The graphical interpretation of $u^\ast$, $u^\dagger$, $v$, $\{\Ucal(v)\}_{\Delta \rho}$, and $f(u)$ is illustrated in Fig.~\ref{fig:U-cost}.
We see that the generated $\Ucal(v) \in \{\Ucal(v)\}_{\Delta \rho}$ illustrated in Fig.~\ref{fig:U-cost}~(b) is more beneficial for $\Hcal$ than Fig.~\ref{fig:U-cost}~(a);
the achievable cost $f(u^\dagger)$ is reduced by the update of $\Ecal$.
We aim to find the best $\Ecal\in\{\Ecal\}_{\Delta\rho}$ in this sense. 

For updating $\Ecal$, we first estimate $u^\ast$ by using some data set $\{v_k, u^\dagger_k\}$, where
$k$ is the discrete time.
Let $E_{\rm L0}$, $E_{\rm L1}$,  $\cdots$, $E_{{\rm L}k}$ be the sequence of the updated $E_{\rm L}$.
We suppose that 
\begin{align} 
	|u_k^\dagger - v_k| < \gamma |E_{{\rm L}k}E_{\rm R}^\top v_k|
	\label{eq:cond1}
\end{align}
holds, 
which implies that $u^\dagger_k$ is located on 
the interior of $\Ucal(v_k)$
as illustrated in Fig.~\ref{fig:U-cost}~(a).
Then, it follows 
that $u^\ast$ is located on the hyperplane described by $E_{{\rm L}k}^\top(u^\ast - u^\dagger_k) = 0$,
which is graphically shown in Fig.~\ref{fig:U-cost}~(a).
The set of the hyperplanes is expressed by the vector form
\begin{align*} 
	[\,E_{\rm L0}\, E_{\rm L1}\, \cdots\, E_{{\rm L}k}\,]^\top u^\ast -  
	[\,E_{\rm L0}^\top u^\dagger_0\, E_{\rm L1}^\top u^\dagger_1\, \cdots\, E_{{\rm L}k}^\top u^\dagger_k\,]^\top = 0.
\end{align*}
If $E_{{\rm ex}k}:= [\,E_{\rm L0}\, E_{\rm L1}\, \cdots\, E_{{\rm L}k}\,]$ is of full row rank, 
we obtain the estimate of $u^\ast$ as 
\begin{align} 
	u^\ast =  (E_{{\rm ex}k}E_{{\rm ex}k}^\top)^{-1}E_{{\rm ex}k} [\,E_{\rm L0}^\top u^\dagger_0\ E_{\rm L1}^\top u^\dagger_1\ \cdots\ E_{{\rm L}k}^\top u^\dagger_k\,]^\top.
	\label{eq:cond2}
\end{align}
The estimate of $u^\ast$ is utilized for updating $\Ecal$.
The algorithm for the update is briefly stated as follows.



\begin{table}[!h]
{\normalsize
\label{alg}
\noindent \hrulefill\\
{{\textit{Algorithm: Updating Expander $\Ecal$}}}\\[-6mm]

\noindent \hrulefill\\[-3mm]
 \begin{algorithmic}[1]
\STATE 	\textit{Initialization}: $E_{{\rm L}k}, E_{{\rm R}k}, \gamma_k$ at $k = 0$
  \REPEAT
	\STATE get data $\{v_k, u^\dagger_k\}$ that satisfies (\ref{eq:cond1})
	\IF {$E_{{\rm ex}k}$ is of full row rank}
		\STATE $E_{{\rm L}k+1} \leftarrow v_k - u^\ast$, where $u^\ast$ is given by (\ref{eq:cond2})
	\ELSE
		\STATE $E_{{\rm L}k+1} \leftarrow E_{{\rm L}k} + \delta$, where $\delta$ is a small perturbation
	\ENDIF
	\STATE find $E_{{\rm R}k+1}$, $\gamma_{k +1}$ maximizing $\gamma_{k+1}|E_{{\rm R}k+1}^\top v_k| $ subject to $\Ecal \in \{\Ecal\}_{\Delta\rho}$
	\RETURN $E_{{\rm L}k+1}, E_{{\rm R}k+1}, \gamma_{k+1}$
	\STATE $k \leftarrow k +1$
  \UNTIL $E_{{\rm L}k}, E_{{\rm R}k}, \gamma_k$ converge
 \end{algorithmic} 
 \hrulefill
}
\end{table}


In the algorithm above, 
it is assumed that $u^\dagger_k$ is available for updating $\Ecal$.
We justify the assumption as follows.
We emphasize that the update can bring benefits only to the {\it decision maker} $\Hcal$, 
not to the  system manager or controller designer who is responsible for the performance of $\Sigma_{\rm HIL}$.  
The benefits for $\Hcal$ can be incentive to disclose some information of $\Hcal$. 
It is thus natural to assume that the result of the decision, denoted by $u^\dagger$, is  disclosed and available for the update of $\Ecal$. 

\section{NUMERICAL EXPERIMENT}

The plant $P$, decision maker $\Hcal$, and controller $\Kcal$ are given as follows.
The plant $P$ is the linear dynamical system described by 
\begin{align*} 
P:\
\left\{\begin{array}{l}
\dot{x} = 
\left[\begin{array}{ccc}
-1 & 0 & 0\\
0 & -2 & 0\\
0 & 0 & -0.5
\end{array}\right]x + 
\left[\begin{array}{cc}
1\\
1\\
1
\end{array}\right]
 w + 
\left[\begin{array}{ccc}
1 & 0 & 0\\
0 & 1 & 0\\
0 & 0 & 1
\end{array}\right] u,\\
y = \left[\begin{array}{ccc}
1 & 1 & 1\\
\end{array}\right]x. 
\end{array}\right.
\end{align*}
The step disturbance is injected to $w$ to drive the plant system.
The transfer matrices from $u$ and $w$ to $y$ are denoted by $P_u(s)$ and $P_w(s)$, $s\in \C$, respectively.  Then, the DC gains of $P_u(s)$ and $P_w(s)$ are given by
\begin{align*} 
	P_u(0) = \left[\begin{array}{ccc}
1 & 0.5 &2
\end{array}\right],\ \ \ 
P_w(0) = 3.5,
\end{align*}
respectively.  
In the decision maker $\Hcal$, the following optimization algorithm is implemented.
\begin{align*} 
\Hcal:\
\left\{\begin{array}{ll}
\min\ &f(u): = 
2 u^\top u - 
\left[\begin{array}{ccc}
1 & 0 &4
\end{array}\right]
u,\\
\mbox{subject to}\ &u \in {\cal U}.
\end{array}\right.
\end{align*}
This optimization model is blind for the design of the controller.
The controller $\Kcal$ is described by
\begin{align*} 
{\cal K}:\
\left\{\begin{array}{l}
v = K_e (r-y + P_u(u)), \\
{\cal U}= v + {\cal E}(v), 
\end{array}\right.
\end{align*}
where $K_e$ is a static system, i.e., it is simply a constant matrix, 
and $P_u$ is the operator representation of $P_u(s)$.
In this section, we demonstrate the design procedure of $K_e$ and $\Ecal$.

The performance criterion for $\Sigma_{\rm HIL}$ is the DC gain, which represents the disturbance suppression performance $|y(t)|$ as $t\rightarrow \infty$ corresponding to the unit step disturbance $w(t)$.
The performance criterion for $\Hcal$ is the value of $f(u^\dagger(t))$.
We aim to minimize $f(u^\dagger(t))$ as $t \rightarrow \infty$ subject to the specification 
$|y(t)| \leq 0.2$ as $t\rightarrow \infty$, denoted by
$|\Sigma_{\rm HIL}|_{\rm dc} \leq 0.2$.

First, 
$K_e$ is designed as
\begin{align*} 
	K_e=\frac{1}{6}\left[\begin{array}{ccc}
			2 & 4 & 1\\
			\end{array}\right],
\end{align*}
which achieves $|y(t)| \rightarrow 0$ as $t\rightarrow \infty$ in the {\it nominal} situation;
in other words, if the expander $\Ecal$ is inactive, the step disturbance $w(t)$ does not propagate to $y(t)$ as $t \rightarrow \infty$.
We see this fact as follows.
Note that  $(1-P_u(s)K_e)P_w(s)$ represents the transfer function of $\Sigma_{\rm HIL}$ when $\Ecal(v) \equiv 0$.  The above $K_e$ guarantees that $(1-P_u(0)K_e)P_w(0) =0$ holds.

Next, the structure of $\Ecal$ is fixed as (\ref{eq:specialE}).
The {\it initial condition} of $E_{\rm L}$ and $E_{\rm R}$ is given by
\begin{align*} 
E_{\rm L0}:=\left[\begin{array}{ccc}
1 & 0 & 0\\
\end{array}\right]^\top,\ \ 
E_{\rm R0} := 
\frac{1}{\sqrt{3}}\left[\begin{array}{ccc}
1& 1&  1\\
\end{array}\right].
\end{align*}
The value of $\gamma$ is determined such that 
the DC gain specification $|\Sigma_{\rm HIL}|_{\rm dc} \leq 0.2$ holds.
The specification is expressed as 
\begin{align*} 
	&|(1-P_u(0)(I_3 + \delta E_{{\rm L}}E_{{\rm R}}^\top)K_e)P_w(0)  |\\
	 &= \delta |P_u(0) E_{{\rm L}}E_{{\rm R}}^\top K_eP_w(0)  |\leq 0.2
\end{align*}
holds for all $\delta \in [\,-\gamma,\gamma\,]$. By maximizing $\gamma$ under the inequality, we obtain  the initial value of $\gamma$  as 
\begin{align*} 
	\gamma_0 = 0.2/|P_u(0) E_{{\rm L0}}E_{{\rm R0}}^\top K_eP_w(0)  | = 0.0848.
\end{align*}
Then, the updating algorithm proposed in Section \ref{sec:app} is applied to update 
$E_{\rm L}$ and $\gamma$, while $E_{\rm R}$ is fixed at $E_{\rm R0}$.

The numerical experiments are performed for the following four cases;
1) no feedback controller is applied, 2) the controller $\Kcal$ is applied without the expander $\Ecal$, 3) the controller $\Kcal$ is applied with fixed $\Ecal$, i.e, $\Ecal$ is composed of $E_{\rm L} = E_{\rm L0}$, $E_{\rm R}=E_{\rm R0}$, and $\gamma= \gamma_0$, and 4) the controller $\Kcal$ is applied with updating $\Ecal$.
The experiment, the time step is fixed at $1$ sec, and the continuous models in $P$ and $\Kcal$ are discretized.  
At each time step, the optimization problem in $\Hcal$ is solved, and the expander $\Ecal$ is updated.

\begin{figure}[t]
 \begin{center}
 \includegraphics[width=0.9\hsize]{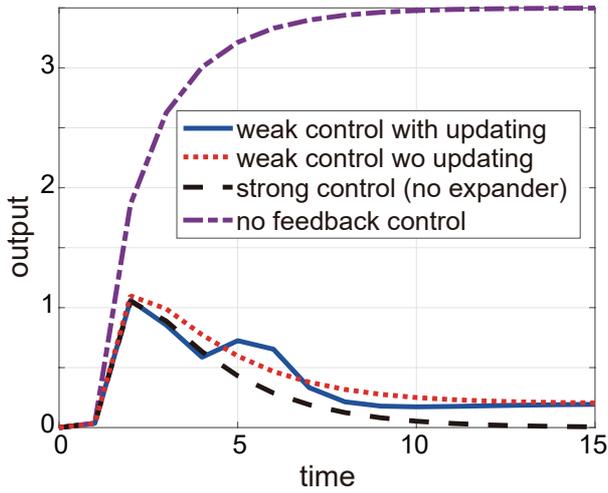}
 \caption{Trajectory of regulated output $y(t)$.
The blue solid, red dotted, and black dashed lines represent the costs achieved by the weak control with the learning mechanism, weak control without any learning mechanism, and strong control, i.e., control without any expander, respectively.  The purple dot-dash line represents the case with no feedback control.}%
 \label{fig:out}
 \end{center}
 \end{figure}

The trajectories $y(t)$ for all cases are illustrated in Fig.~\ref{fig:out}.
We see that the feedback control effectively suppresses the disturbance effects in $y(t)$.
The control in Case 2 results in the best performance, while
the weak control in Cases 3 and 4 satisfies the specification, $|y(t)|\leq 0.2$ at a large $t$.

\begin{figure}[t]
 \begin{center} 
 \includegraphics[width=0.9\hsize]{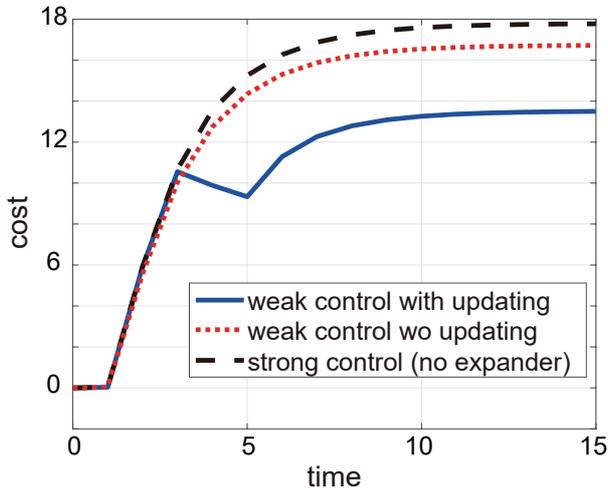}
 \caption{Trajectory of cost $f(u^\dagger(t))$ for $\Hcal$.  
The blue solid, red dotted, and black dashed lines represent the costs achieved by the weak control with the learning mechanism, weak control without any learning mechanism, and strong control, i.e., control without any expander, respectively.}%
 \label{fig:cost}
 \end{center}
 \end{figure}

The values of the cost $f(u^\dagger(t))$ for all cases are illustrated in Fig.~\ref{fig:cost}.
We see that the costs achieved by the weak control in Cases 3 and 4 are smaller than that by the control in Case 2. This demonstrates that the expander $\Ecal$ brings smaller costs for decision makers $\Hcal$.
Furthermore, we compare Cases 3 and 4 to show the effectiveness of the updating algorithm.  The weak control with updating $\Ecal$ in Case 4 contributes to reducing the cost  compared with no updating case in Case 3 as illustrated in Fig.~\ref{fig:cost}.  
It should be emphasized that Case 4 further reduces the cost while keeping the same DC gain performance in $\Sigma_{\rm HIL}$ as illustrated in Fig.~\ref{fig:out}.


\section{CONCLUSION}
In this letter, we proposed a framework for {\it weak control}  for human-in-the-loop systems.  
In this framework, a signal {\it expander} is embedded in the controller and 
generates candidate control actions with some DOF.   
The DOF allows the human decision-makers to pursue their own aims, while guaranteeing the stability and the specified performance in the overall control system.  
A simple algorithm of  updating the expander was also given, which was beneficial to human decision makers.

There are a variety of future works for the weak control; 
more sophisticated algorithms of updating the expander can be derived under practical problem setting, and the weak control can be applied to the demand response for power grids \cite{Palensky_11}. 


\bibliographystyle{IEEEtran}
\bibliography{Ref_minoue}

\begin{thebibliography}{10}
\providecommand{\url}[1]{#1}
\csname url@rmstyle\endcsname
\providecommand{\newblock}{\relax}
\providecommand{\bibinfo}[2]{#2}
\providecommand\BIBentrySTDinterwordspacing{\spaceskip=0pt\relax}
\providecommand\BIBentryALTinterwordstretchfactor{4}
\providecommand\BIBentryALTinterwordspacing{\spaceskip=\fontdimen2\font plus
\BIBentryALTinterwordstretchfactor\fontdimen3\font minus
  \fontdimen4\font\relax}
\providecommand\BIBforeignlanguage[2]{{%
\expandafter\ifx\csname l@#1\endcsname\relax
\typeout{** WARNING: IEEEtran.bst: No hyphenation pattern has been}%
\typeout{** loaded for the language `#1'. Using the pattern for}%
\typeout{** the default language instead.}%
\else
\language=\csname l@#1\endcsname
\fi
#2}}

\bibitem{Whitney_69}
D.~E. Whitney, ``State space models of remote manipulation tasks,'' \emph{IEEE
  Transactions on Automatic Control}, vol.~14, no.~6, pp. 617--623, 1969.

\bibitem{Mcruer_80}
D.~McRuer, ``Human dynamics in man-machine systems,'' \emph{Automatica},
  vol.~16, no.~3, pp. 237--253, 1980.

\bibitem{Sheridan_89}
T.~B. Sheridan, ``Telerobotics,'' \emph{Automatica}, vol.~25, no.~4, pp.
  487--507, 1989.

\bibitem{Hokayem_06}
P.~F. Hokayem and M.~W. Spong, ``Bilateral teleoperation: {A}n historical
  survey,'' \emph{Automatica}, vol.~42, no.~12, pp. 2035--2057, 2006.

\bibitem{Goodrich_08}
M.~A. Goodrich and A.~C. Schultz, ``Human--robot interaction: {A} survey,''
  \emph{Foundations and Trends{\textregistered} in Human--Computer
  Interaction}, vol.~1, no.~3, pp. 203--275, 2008.

\bibitem{Cao_08}
M.~Cao, A.~Stewart, and N.~E. Leonard, ``Integrating human and robot
  decision-making dynamics with feedback: Models and convergence analysis,'' in
  \emph{Proceedings of the 47th IEEE Conference on Decision and Control}, 2008,
  pp. 1127--1132.

\bibitem{Chipalkatty_13}
R.~Chipalkatty, G.~Droge, and M.~B. Egerstedt, ``Less is more: Mixed-initiative
  model-predictive control with human inputs,'' \emph{IEEE Transactions on
  Robotics}, vol.~29, no.~3, pp. 695--703, 2013.

\bibitem{Lam_14}
C.-P. Lam and S.~S. Sastry, ``A {POMDP} framework for human-in-the-loop
  system,'' in \emph{Proceedings of the 53rd IEEE Conference on Decision and
  Control}.\hskip 1em plus 0.5em minus 0.4em\relax IEEE, 2014, pp. 6031--6036.

\bibitem{Hatanaka_CDC15}
T.~Hatanaka, N.~Chopra, and M.~Fujita, ``Passivity-based bilateral
  human-swarm-interactions for cooperative robotic networks and human passivity
  analysis,'' in \emph{Proceedings of the 54th IEEE Conference on Decision and
  Control}, 2015, pp. 1033--1039.

\bibitem{Music_17}
S.~Musi{\'c} and S.~Hirche, ``Control sharing in human-robot team
  interaction,'' \emph{Annual Reviews in Control}, 2017.

\bibitem{Palensky_11}
P.~Palensky and D.~Dietrich, ``Demand side management: {D}emand response,
  intelligent energy systems, and smart loads,'' \emph{IEEE Transactions on
  Industrial Informatics}, vol.~7, no.~3, pp. 381--388, 2011.

\bibitem{Wickens_97}
C.~D. Wickens, A.~S. Mavor, and J.~P. McGee, \emph{Flight to the Future: Human
  Factors in Air Traffic Control}.\hskip 1em plus 0.5em minus 0.4em\relax
  National Academies Press, 1997.

\bibitem{Barfield_97}
W.~Barfield and T.~A. Dingus, \emph{Human Factors in Intelligent Transportation
  Systems}.\hskip 1em plus 0.5em minus 0.4em\relax Psychology Press, 1997.

\bibitem{Maestre_14}
J.~M. Maestre, P.-J. van Overloop, M.~Hashemy, A.~Sadowska, and E.~F. Camacho,
  ``Human in the loop model predictive control: An irrigation canal case
  study,'' in \emph{Proceedings of the 53rd IEEE Conference on Decision and
  Control}, 2014, pp. 4881--4886.

\bibitem{Van_15}
P.~Van~Overloop, J.~Maestre, A.~D. Sadowska, E.~F. Camacho, and B.~De~Schutter,
  ``Human-in-the-loop model predictive control of an irrigation canal,''
  \emph{IEEE Control Systems Magazine}, vol.~35, no.~4, pp. 19--29, 2015.

\bibitem{Feng_16}
L.~Feng, C.~Wiltsche, L.~Humphrey, and U.~Topcu, ``Synthesis of
  human-in-the-loop control protocols for autonomous systems,'' \emph{IEEE
  Transactions on Automation Science and Engineering}, vol.~13, no.~2, pp.
  450--462, 2016.

\bibitem{Eichler_17}
A.~Eichler, G.~Darivianakis, and J.~Lygeros, ``Humans in the loop: {A}
  stochastic predictive approach to building energy management in the presence
  of unpredictable users,'' \emph{IFAC-PapersOnLine}, vol.~50, no.~1, pp.
  14\,471--14\,476, 2017.

\bibitem{Garcia_82}
C.~E. Garcia and M.~Morari, ``Internal model control: {A} unifying review and
  some new results,'' \emph{Industrial \& Engineering Chemistry Process Design
  and Development}, vol.~21, no.~2, pp. 308--323, 1982.

\bibitem{Zhou_96}
K.~Zhou, J.~C. Doyle, K.~Glover, \emph{et~al.}, \emph{Robust and Optimal
  Control}.\hskip 1em plus 0.5em minus 0.4em\relax Prentice Hall, 1996.

\bibitem{LCSS_18}
M.~Inoue, ``Persistence in control systems,'' \emph{IEEE Control Systems
  Letters}, vol.~2, no.~3, pp. 387--392, 2018.

\end{thebibliography}
\noindent
\end{document}